# Some paradoxes in the Tilman model, how to avoid or accept them.

*Joost H.J. van Opheusden*[1,*], *Lia Hemerik*[1]

[1)] *Biometris, Wageningen University, Wageningen, The Netherlands.*

[*)] *corresponding author: Joost H.J. van Opheusden; e-mail* joost.vanopheusden@gmail.com

## Abstract

In making mathematical models for biological systems one expects to grasp the biology and be able to use one's intuition to predict the outcome. This paper is about the discrepancy between what is expected and what is the outcome of the analysis of a differential equation model, a so-called paradox. The Tilman model of consumers competing for resources has some aspects that may appear counterintuitive. Within the standard Tilman model for a single consumer and a single resource, when a very efficient consumer rapidly eats all available food, the resource density becomes zero, but if there is no food, how can the consumer survive? The paradox can be resolved by realising that in the short term indeed rapid consumption may lead to starvation and a decline in the consumer population, but in the long term a small resource and consumer density remain. A single consumer living on two essential nutrients leaves the density of the non-limiting nutrient above its critical level, so what is done with the extra food? We explain that the extra food is not consumed, just because it is available, but that in fact the additional food is not consumed at all, because the consumer has no use for it. For a model with two consumer species competing for a single nutrient one of the consumers will eventually disappear, even if the food economics of the surviving one is worse and there is still much food wasted. Both are inherent aspects of the model, and the paradox can be avoided if the difference between the consumer species is small, in which case it will take very long to reach equilibrium.

We argue that in general an extended stability analysis, in which not only the asymptotically stable state is considered, but also the unstable steady states and all time scales involved in the transient dynamics, gives more insight in the dynamics involved and thus can help in avoiding apparent contradictions in ecological models, or accepting them.

## 1. Introduction

***Paradox*** noun, UK /ˈpær.ə.dɒks/, US /ˈper.ə.dɑːks/: *a situation or statement that seems impossible or is difficult to understand because it contains two opposite facts or characteristics*. (Cambridge Dictionary Online)

### 1.1. Paradoxes

This paper is not about paradoxes. Binary logic will have it that the previous statement is either true or false. If it is true, the title is wrong, if it is false, then the introduction is wrong. A paradox; this paper cannot be about paradoxes and at the same time not be about paradoxes. In this case the solution is that the statement is wrong: this paper *is* about paradoxes. At the same time the statement is right (or at least not entirely wrong); the paper deals with some counterintuitive aspects of biological models, which can be seen as paradoxical. Philosophers have used the idea of a paradox since ancient times in order to stimulate a critical discussion about a subject, to sharpen the mind. These paradoxes were about logical contradictions, statements that can be perceived as both true and false. The discussion then is about the very concept of logic, a discussion that is still very much ongoing [*for a general introduction into these paradoxes and a classification see for instance* Stanford Encyclopedia of Philosophy, Internet Encyclopedia of Philosophy, or Sainsbury (2009)]. Here we apply this same approach to discuss some issues that may arise when using simple models for ecological systems. The paradoxical situations arise as a conflict between the results of the model and the expected behaviour of it, based on intuitions derived from general observations of the kind of systems that are modelled. If the model does not perform as expected, what is wrong? Maybe it is the model, but it is well possible that it is the expectation, or the interpretation of the model, or something else. By formulating several paradoxes, and analyzing these in detail, we want to discuss and elucidate some aspects of simple ecological systems that still often lead to confusion, especially for people who are not (yet) used to modelling. For the sake of argument we will not stop at misleading the reader first in formulating an alleged paradox, which, admittedly, may cause frustration, but it comes with the territory and obviously we hope that in the end it adds to the understanding. In particular we discuss some paradoxes of the Tilman model of competition for resources: a single consumer eating so rapidly that no food is left to survive, a consumer eating very rapidly but not using the food to build a sizeable population, still outperforming one that does, and a consumer seemingly eating one food out of luxury, while not actually needing it. As an introductory example we start with a very simple textbook model in ecology, not the Tilman model, to show what we mean.

### 1.2. A simple example

A farmer grows lettuce, which is eaten by snails. In a very simple model the ecosystem represented by the biomass $L(t)$ of lettuce and snails $S(t)$ is modelled as

$$\begin{cases} \frac{\mathrm{d}L(t)}{\mathrm{d}t} = & aL(t)\left(1-\frac{L(t)}{x}\right)-bL(t)S(t) \\ \frac{\mathrm{d}S(t)}{\mathrm{d}t} = & -cS(t)+dL(t)S(t) \end{cases} \qquad (0)$$

with $a$, $b$, $c$, and $d$ positive constants. In the absence of snails, the lettuce grows with an initial rate $a$, and a carrying capacity $x$. In the absence of lettuce, the snails die at a rate $c$. If both the lettuce and the snails are present the snails eat the lettuce at a rate proportional with both densities and proportionality constant $b$, with the lettuce biomass turned into snail biomass, at a yield ratio $d/b$, with a constant $d$. There are three equilibria, (1) no lettuce and no snails, (2) lettuce at carrying capacity and no snails, and (3) the equilibrium with positive densities of both lettuce and snails with $L_S = c/d$ and $S_S = (1 - L_S/x)\ a/b$ for the lettuce and snails respectively. We first take a look at the equilibrium with both populations present. We assume the carrying capacity for the lettuce is much larger than the equilibrium density ($x >> L_S$). In equilibrium any increase in lettuce biomass is eaten away by the snails, and any increase in snail biomass is nullified by snails dying. In equilibrium the farmer is producing dead snails rather than lettuce.

Assuming this is not what the farmer had in mind, two agricultural options are available: (I) kill off the snails, or (II) speed up the growth of the lettuce. The first option amounts to an increase in *c*, the mortality rate of the snails. The second option amounts to an increase in *a*, the growth rate of the lettuce. The parameters *b* and *d* describe the interaction between the snails and the lettuce, which is no business of the farmer. Because of the (changed) interference of the farmer with the ecosystem, the equilibrium shifts.

***Paradox 1:*** Now we find a paradox. If the mortality rate *c* of the snails is increased, it is not the snail density at equilibrium that is heavily affected, but rather the lettuce density that increases. If on the other hand the growth rate *a* of the lettuce is increased, the equilibrium lettuce density is unaltered, instead the snail density goes up. ***This is counterintuitive, we expect killing off snails to bring the snail density down. Similarly boosting lettuce growth should increase lettuce density.*** Indeed that is exactly what happens, right after the parameter is changed, but after a while the effect is counteracted by the response from the interacting snail-lettuce system.

If we investigate the dynamics for increased snail removal in detail (numerically integrating the equations using an forward Euler method with sufficiently small time step), we see (fig 1) that an initial decrease in snails leads to an increase in lettuce, which does not go unnoticed by the snails. Their density increases as well, and the system goes to a number of cycles of increasing and decreasing snail and lettuce density, until an equilibrium is reached in which all lettuce growth is converted into dead snails. More dead snails, because their mortality rate is much higher.

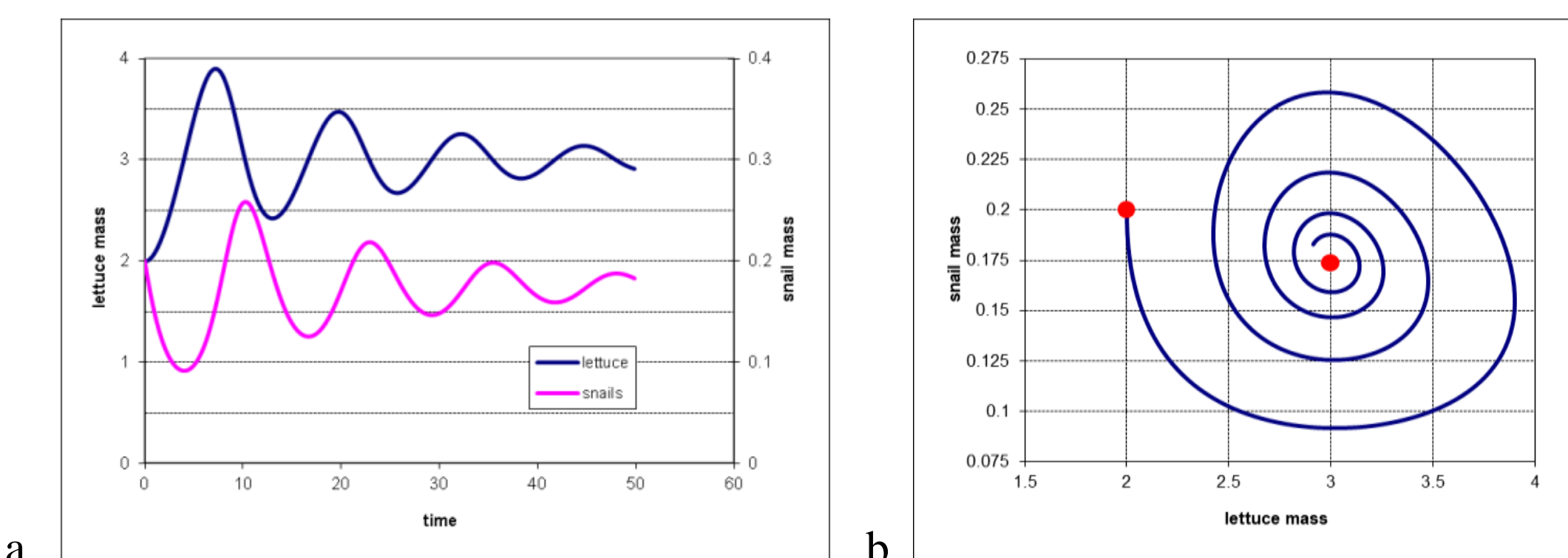

fig. 1. Effect of an increase in death rate *c* of the snails in the snail-lettuce model. For parameter values see the appendix. On the short term the snail mass *S* (cyan line in fig a) goes down as expected, while the lettuce mass *L* goes up (blue line). On the long term the lettuce mass does go up less, the snail mass goes down only slightly. The phase diagram (fig b) shows how the system spirals from the original equilibrium to the new one (red dots), with an increased lettuce mass (horizontal axis) and almost the same snail mass (vertical axis).

More fertilizer initially leads to an increase in lettuce (fig 2), and again the snails seize the opportunity. Lettuce density goes down and up again until an equilibrium value is reached with the same amount of lettuce, more snails, and hence an elevated production of dead snails.

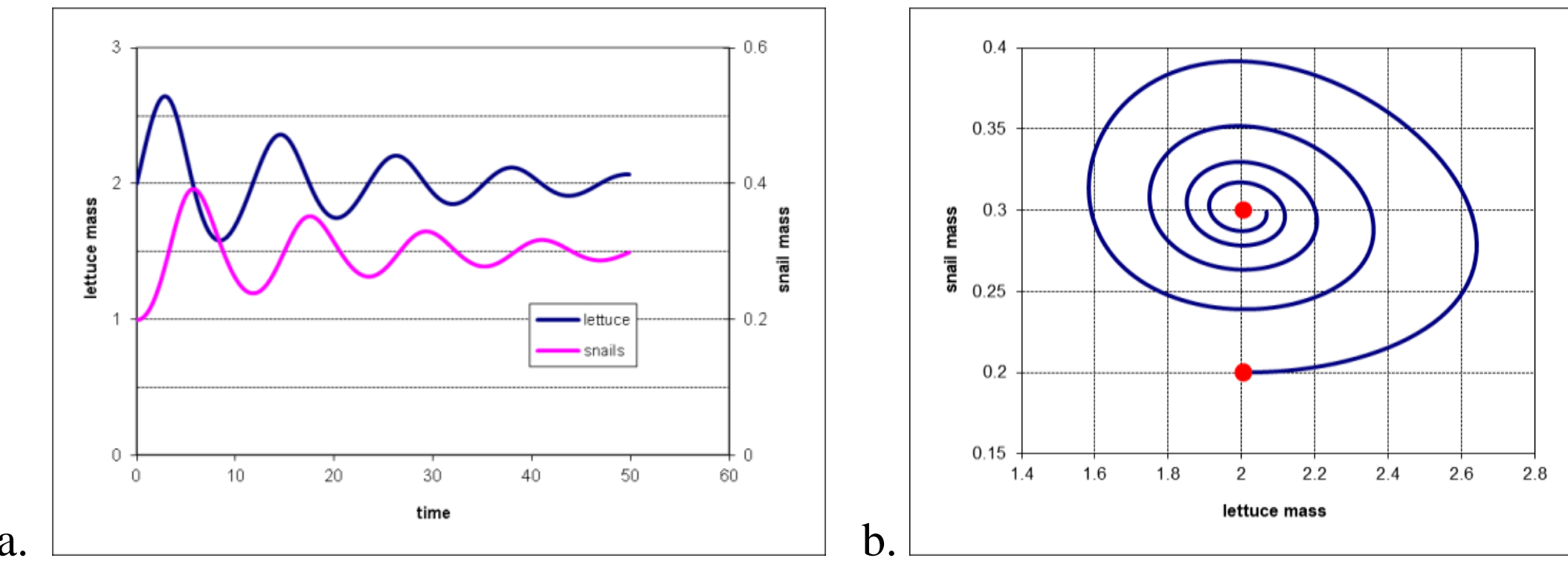

fig 2. Effect of an increase in growth rate $a$ of the lettuce. On the short term the lettuce mass $L$ (blue line in fig a) indeed goes up, but so does the snail mass $S$ (cyan line). On the long term the lettuce mass returns to the original equilibrium value, the snail mass goes up. The phase diagram (fig b) shows how the system moves in cycles from the original equilibrium value to the new one, with the same lettuce mass and an increased snail mass.

In fact there is no contradiction, it only looks like there is one, so we have a paradox indeed. ***The system does react instantaneously as we would expect it to do, but in the long run it does quite the opposite, and the effect can be explained and understood by looking at the detailed dynamics of the system.*** There is not just one equilibrium (as already mentioned above); in fact the system has three (red dots in fig 3b). Starting near the saddle point at zero snail and lettuce density, the system first moves towards a saddle point with only lettuce. Because there are still a few snails, they eventually start multiplying and eating a sizeable portion of the lettuce, so the system moves away from the second saddle point as well, starting a series of loops with alternating high and low snail and lettuce density that will ultimately bring it to a stable coexistence equilibrium. This is what the mathematics tells us, but any sensible farmer growing lettuce will not wait for this, but harvest the lettuce right before the snails claim their share, close to the second equilibrium, the saddle point with zero snail and high lettuce density. Better avoid the snails than attack them.

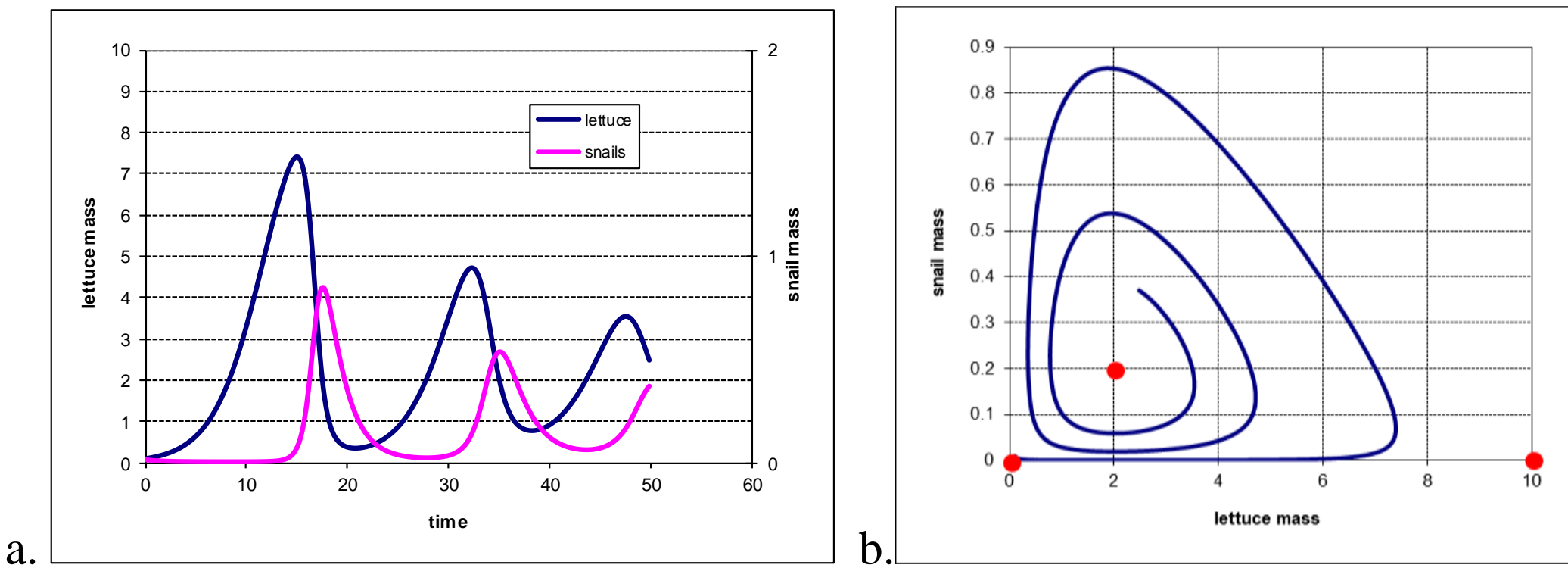

fig 3. Dynamics of the snail-lettuce system starting at low lettuce and snail mass. The lettuce mass grows rapidly (a), while the snail mass remains very low. Once the snails start multiplying seriously, they eat away most of the lettuce, snails die, and the cycle repeats. Eventually the system will spiral towards the stable coexistence point (phase plot b).

The time scale at which the lettuce is harvested by the farmer is much shorter than that at which the system relaxes towards the final equilibrium. So, the relevant time scale in this case is the short one, the relevant equilibrium an unstable one. The cause of the paradox, from the perspective of the farmer growing lettuce, is looking at the wrong time scale. On the short time scale, after sowing the lettuce with no snails around, the model behaves as expected. Adding fertilizer does increase the growth rate of the lettuce. When snails enter the field, killing them off does prevent them from eating the lettuce. Similarly if a few snails are present, it helps to prevent them from multiplying. From the perspective of a farmer growing lettuce, the stable final equilibrium is irrelevant, or is something that should be avoided. For a farmer growing snails, a different perspective gives a different picture, of course, and similarly for the snails, the lettuce and the theoretical biologist.

This is an example of how a relatively simple mathematical model, and a simple model analysis, can lead to a paradox. Note that the paradox will not be apparent to everyone; an experienced scientist will probably have encountered the phenomenon earlier and doesn't see the observed model behaviour as counterintuitive, but rather as familiar. For those readers the approach may serve as a suggestion on how to discuss such perceived contradictions with less experienced readers.

### 1.3. Outline of the paper

We will show how similar paradoxes can occur in a simple system of consumers competing for resources, the Tilman model (Tilman, 1977; Tilman, 1982). The Tilman model is substantially more complicated than the simple toy model treated above. Indeed we have used the toy model as a somewhat peculiar type of introduction, in order to provide the right perspective for biologists. The main lesson is to never put blind trust in your biological intuition when it comes to assessing equilibrium dynamics in models of ecological systems, the main question is what one would want to do with the resulting paradoxes; accept, or resolve them. This will be the focus of the discussion below. For a single consumer of a single resource, which we treat in paragraph 2.1, the maximum consumer density is obtained if the consumers eat away all of the resource, but if there is no food, how can there be consumers at all? For two consumer species competing for a single resource we show in paragraph 2.2 it may happen that a consumer with a bad food efficiency, when invading a system with a consumer with a (much) better efficiency, still outperforms the latter and takes over the system. Why isn't there a bonus on food efficiency, and why does the one with the better efficiency become extinct when there seems to be still plenty of food around? In paragraph 2.3 we show for a single consumer depending on two resources, that it seems as if the consumer is always eating more than is needed of one of the two resources, without any positive effect on its survival. Why would that consumer bother eating an excessive amount of that resource?

## 2. The Tilman model

Competition is an important concept in how we think about ecosystems. Whether on an individual level or on that of a population or species, the idea is that certain types of behaviour provide an advantage as they promote survival or procreation and form a driving force of the development of the system. In the Tilman model, which was originally formulated for algae and later extended towards plants, different consumer populations compete for resources, not directly in the sense that one consumer species actively fights with another about access to resources, but indirectly, for instance by faster consumption of the resource, or larger growth rates at the same consumption rate. While in reality a wide variety of behavioural and physiological aspects may come into play, in the models we treat there are just a few parameters that describe the interaction between the consumer and the resource and the consumer's reaction on intake of resources. We will use the term 'foraging' to label those parameters that can be associated with the intake of resources. In the models we compare the competition between two different consumer species and look for the one with the combination of parameter values that gives that consumer an advantage over the other. These describe the strategy that in the long run makes this consumer prevail. Note that the term strategy, like competition, suggests intentional behaviour; in the models there is no reason to assume there is, but the effect is the same. Note also that the notion of advantage is to be understood in this context, it is not a game in which the "better" consumer wins. Though it often helps to describe the working of the biological system (and that of the model) in such terms, it may also add to the confusion and be the source of the conflict between model result and intuition.

In an earlier paper (Van Opheusden, 2015) we have extensively described the competition model as introduced by Tilman, and performed a stability analysis. We summarize the discussion briefly. In the model there are consumers that inhabit a given ecosystem, and resources that the consumers need for their maintenance and growth. The model describes the development of the consumer and resource densities in time. There is no direct interaction between the consumers, nor between the resources. The dynamics of the densities is described by a system of coupled ordinary differential equations. The form for a system with a single consumer population with a biomass density $B(t)$ and a single resource biomass density $R(t)$ is

$$\begin{cases} \frac{\mathrm{d}B(t)}{\mathrm{d}t} = f_{\mathrm{B}}(R(t))B(t) - m_{\mathrm{B}}B(t) \\ \frac{\mathrm{d}R(t)}{\mathrm{d}t} = a_{\mathrm{R}}(s_{\mathrm{R}} - R(t)) - q_{\mathrm{RB}}f_{\mathrm{B}}(R(t))B(t) \end{cases}, \tag{1}$$

The parameter $m_{\mathrm{B}}$ is the mortality rate of the consumer. The growth rate of the consumer is a function $f_{\mathrm{B}}(R(t))$ of the resource density only. We use a Holling type II functional response,

$$f_{\mathrm{B}}(R(t)) = f_{\mathrm{Bm}} \frac{R(t)}{R(t)+k_{\mathrm{RB}}}, \tag{2}$$

with maximum rate $f_{\mathrm{Bm}}$ and half saturation constant $k_{\mathrm{RB}}$. Decrease in resource density as a result of feeding leads to a proportional increase of the consumer density, with a fixed and constant conversion factor $q_{\mathrm{RB}}$. In the absence of consumers the resource density is described by a so-called chemostat equation, with a dilution rate $a_{\mathrm{R}}$ and a supply density $s_{\mathrm{R}}$. Additional consumer species and resources can be added to the model along the same lines, each consumer having a growth rate depending only on the resource densities and the depletion of the resources simply equal to the sum of the consumption by all consumers.

The system has two stationary states, a trivial one in which the consumer is fully absent

$$R = s_{\mathrm{R}}, \;\; B = 0 \tag{3}$$

and one in which the consumer coexists with the resource at a finite density

$$R = R^* = \frac{k_{\mathrm{RB}}m_{\mathrm{B}}}{f_{\mathrm{Bm}}-m_{\mathrm{B}}}, \;\; B = B^* = \frac{a_{\mathrm{R}}(s_{\mathrm{R}}-R*)}{q_{\mathrm{RB}}m_{\mathrm{B}}} \tag{4}$$

Even for a single consumer and a single resource the model contains quite a number of parameters, one for the consumer ($m_{\mathrm{B}}$), two for the resource ($a_{\mathrm{R}}, s_{\mathrm{R}}$), and three for their interaction ($f_{\mathrm{B}m}, k_{\mathrm{RB}}, q_{\mathrm{RB}}$). All ecological effects within a complex ecosystem must enter the model through these parameters.

### 2.1. A single consumer and a single nutrient

In the absence of consumers the dynamics of abiotic resources is given by the chemostat equation (Tilman, 1977):

$$\frac{\mathrm{d}R(t)}{\mathrm{d}t} = a_{\mathrm{R}}(s_{\mathrm{R}} - R(t)). \tag{5}$$

***Paradox 2:*** There is a paradoxical aspect to this equation. The equation describes the working of a chemical device called a chemostat, used in controlled chemical experiments and production processes. It consists of a reaction tank of volume $V$, containing a well-mixed solution of a nutrient at a varying concentration $R(t)$. The reaction tank is supplied with a nutrient solution at a constant rate $F$ from a feeder tank, with a constant concentration $s_{\mathrm{R}}$ , and there is an outflow at the same rate, keeping the volume of the reaction tank fixed. The ratio $V/F$ is the average residence time of a nutrient molecule inside the reaction tank, defining a time scale of the process. Inversely $a_{\mathrm{R}} = F/V$ is called the dilution rate of the chemostat. The inflow into the reaction vessel is fixed at $a_{\mathrm{R}}s_{\mathrm{R}}$, the outflow $a_{\mathrm{R}}R(t)$ depends on the resource density. A chemostat equation will also apply for instance to a lake with nutrients delivered and removed through inflowing and outflowing streams.

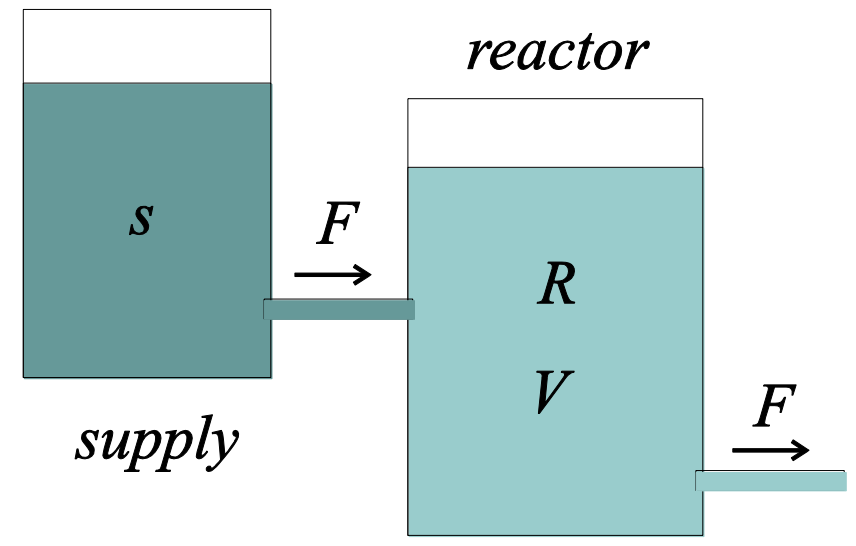

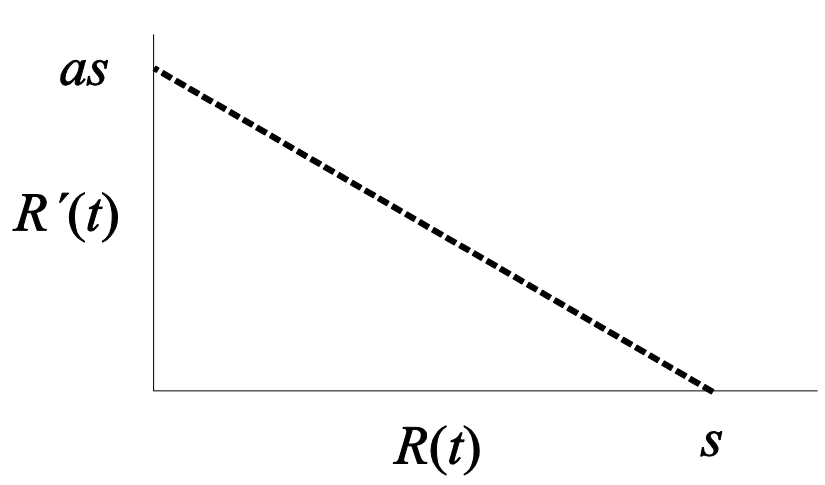

Figure 4. Schematics of a chemostat. Nutrient flows at a rate *F* from a supply vessel with a constant concentration *s* to a reactor with a constant volume *V*.

Figure 5. Phase plot of a chemostat. The net nutrient inflow $R'$ decreases linearly with concentration *R* and is maximal at zero concentration.

Now what is the paradox? If there are no consumers, eventually a stable resource density $s_{\mathrm{R}}$ is reached, inflow matches outflow and all nutrients pass the system untouched. If there are consumers, part of the inflow is used, leading to an extra term in the equation. The resource density drops and the outflow is reduced proportionally. In the optimal situation, from the consumer perspective, consumers manage to eat all inflowing nutrients, nothing goes to waste, and the net outflow is zero. The nutrient density then is zero too. In equilibrium there is a finite positive nutrient density $R^*$ (eqn 4), and what the consumers cannot manage to eat flows out of the system. Consumers can use a different foraging strategy to increase the amount eaten. Regardless of whether improved consumer behaviour leads to a decrease in mortality ($m_{\mathrm{B}}$) or the half saturation constant ($k_{\mathrm{RB}}$), or increase in the intake rate ($f_{\mathrm{B}m}$), the equilibrium density $R^*$ goes down. As that implies less food leaves the system untouched, such a strategy can be seen as more efficient. For the chemostat equation there is a negative relation between the density of the nutrient and the net resource supply rate (inflow rate minus outflow rate, Fig. 5). This is the basis of the paradox; ***in this limiting case an infinitely efficient consumer eats all nutrients that flow into the system, leaving the nutrient concentration in the system at zero, and if there are no nutrients to feed on left, how can the consumer survive?***

A better look at the model shows that in the limit of $k_{\mathrm{RB}}$ becoming zero, indeed the equilibrium nutrient density becomes zero, and the equilibrium consumer density $B^*$ goes to its maximum. The full inflow of nutrients is converted into consumer biomass at the maximal rate, independent of the nutrient concentration (eqn 2) and in that sense the consumer behaviour is optimal. The model also indicates we shouldn't worry about the consumers having no food, because the maximal intake is independent of the food concentration, but depends on the inflow rate.

The question is whether this provides a realistic model, can we set $k_{\mathrm{RB}}$ to zero and leave all other parameters the same? If food becomes scarce, because it is eaten rapidly by the consumers, finding the last remaining nutrients will come at a cost. If the mortality goes up as a result of the increased effort needed to forage for the food, the net effect on the biomass can even be negative. However, these mechanisms are not represented in the model. There is nothing inherently wrong with the model, it is simply an extreme case, the mathematics of which can be understood by considering the limit of low $k_{\mathrm{RB}}$ and it makes biological sense (see below) that the inflow rate and not the food concentration determines the biomass density of the consumer. The food density is zero, or extremely low, because everything is rapidly consumed compared to the dilution rate. Optimized consumer behaviour will not focus at reducing the nutrient spill to zero, that is a human perspective. A consumer population in this model system may adopt a foraging strategy that maximizes its biomass, and if some food is spilled in the process, so be it, there is still plenty coming in (also see subsection 2.2). ***It is the net resource supply rate that is relevant for sustaining the population, not the (equilibrium) concentration.***

Moreover, there was nothing wrong with our intuition, if the consumers rapidly eat the food, there will not enough be left to survive on and they will die. What will happen depends on the rate of the nutrient and consumer dynamics, which in the model is given by the parameters $a_{\mathrm{R}}$ and $f_{\mathrm{Bm}}$ respectively. If the rates are comparable, the system will move gradually to equilibrium, but if the rate of the consumer dynamics is much larger than that of the nutrient (implying the associated time scale is much shorter) the consumer density can overshoot the stable one. In fig 6 this is shown for parameter values $a_{\mathrm{R}} = 0.05$, $m_{\mathrm{B}} = 1$, $s_{\mathrm{R}} = 1$, $f_{\mathrm{Bm}} = 2.5$, $k_{\mathrm{RB}} = 0.2$, $q_{\mathrm{RB}} = 1$. The orbit in the phase plane starts near the saddle point where the nutrient density is at the supply level. The dilution rate is small compared to the consumption rate and consequently the consumers eat all the available food, as expected. Once the food is gone, the consumers start dying, consumption drops, gradually food levels increase, and the cycle starts again. The orbit spirals towards the equilibrium, which is a stable vortex. We have $R^* = 0.13$, so 87% of the available food supply is used. The maximum value of the consumer biomass density reached during the process is about 0.40, much higher than the equilibrium density $B^* = 0.043$. Far away from equilibrium the system behaves as expected, but eventually it is a subtle balance that determines the fate of the consumer. In a system where the food density reaches a very low level, all consumers may die because of fluctuations in either the food recovery or consumer extinction, but these fluctuations are not included in the model. Thus this is another way out of the paradox; in reality consumers voraciously eating away the resources can become extinct, but the mechanism that produces this effect is not part of the model.

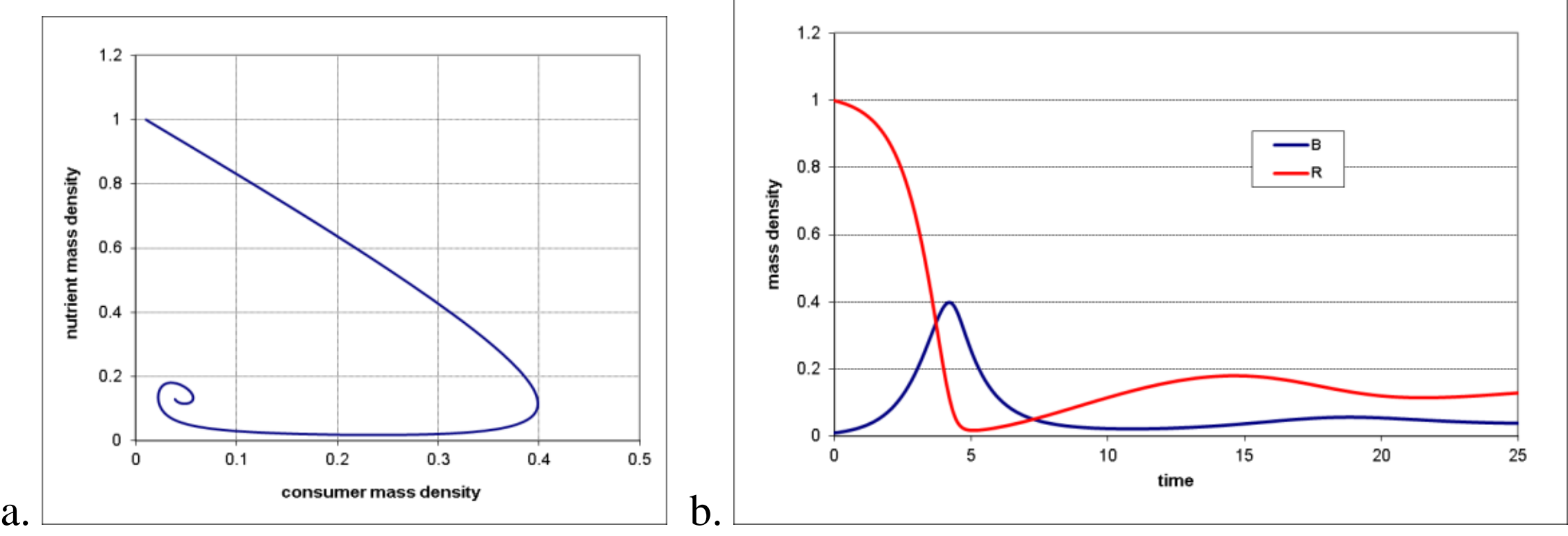

fig 6. Dynamics of the nutrient and consumer dynamics for a system according eqn 1 with parameters $a_{\mathrm{R}} = 0.05$, $m_{\mathrm{B}} = 1$, $s_{\mathrm{R}} = 1$, $f_{\mathrm{Bm}} = 2.5$, $k_{\mathrm{RB}} = 0.2$, $q_{\mathrm{RB}} = 1$. The low value of the half saturation constant $k_{\mathrm{RB}}$ indicates efficient consumption. Initially there is plenty of food, so the consumption rate is maximal and (almost) independent of the nutrient concentration.

Consequently consumer biomass density increases exponentially (b), while nutrient mass decreases likewise, because of the low dilution rate. When the food runs out, consumer biomass drops exponentially, until the food level is high enough to turn things around. Eventually the system reaches the stable equilibrium (a) at which most of the food supply is consumed. Because the food supply rate is low, the equilibrium consumer density is low too.

**2.2. Two consumers competing for a single resource.**

Above we have looked at some aspects of the simplest Tilman model, with one consumer and one nutrient. Whether or not the consumer can survive on the available food supply depends on its properties. As expressed in the model parameters: $R^* < s_{\mathrm{R}}$, or $f_{\mathrm{B}m} > m_{\mathrm{B}}(1 + k_{\mathrm{RB}}/s_{\mathrm{R}})$, otherwise the food supply is not enough to maintain a stable, positive, population density. The other parameters further determine the biomass density the consumer can reach at equilibrium. We now ask the question whether a consumer can do better than just (barely) survive. Given fluctuations that are present in any real ecosystem, a consumer can increase its chances of survival by increasing the stationary consumer biomass density $B^*$. When it comes to foraging, all parameters except the parameters $\mathrm{a}_{\mathrm{R}}$ and $s_{\mathrm{R}}$ that define the resource dynamics in absence of the consumers are open for improvement: in order to have a higher $B^*$, according to eqn (4), either the maximal growth rate $f_{\mathrm{Bm}}$ can be increased, or the half saturation constant $k_{\mathrm{RB}}$ can be lowered, or the conversion factor $q_{\mathrm{RB}}$ (the inverse of the yield factor) can be decreased. These parameters directly affect the consumption. The mortality must not go up too much, otherwise all effort is in vain. In most cases better behaviour will be a combination of all four effects, so that the net result is positive. Let us assume a consumer develops a behavioural pattern that reduces the energy involved in acquiring the resource. Lower energy losses in acquiring the food make that an individual consumer can reach the same growth rate with a smaller amount of food, or a higher growth rate with the same amount. The question is what would be the better strategy. Since the Tilman model itself is intended to describe competition, we use the same model, but add another consumer species, with the same mortality, but with one of the other parameters slightly different. Again we will find a paradox.

The model of two consumers ($A$ and $B$) and a single resource is

$$\begin{cases} \frac{\mathrm{d}A(t)}{\mathrm{d}t} = f_{\mathrm{Am}} \frac{R(t)}{R(t)+k_{\mathrm{RA}}} A(t) - m_{\mathrm{A}} A(t) \\ \frac{\mathrm{d}B(t)}{\mathrm{d}t} = f_{\mathrm{Bm}} \frac{R(t)}{R(t)+k_{\mathrm{RB}}} B(t) - m_{\mathrm{B}} B(t) \\ \frac{\mathrm{d}R(t)}{\mathrm{d}t} = a_{\mathrm{R}}(s_{\mathrm{R}} - R(t)) - q_{\mathrm{RA}} f_{\mathrm{Am}} \frac{R(t)}{R(t)+k_{\mathrm{RA}}} A(t) - q_{\mathrm{RB}} f_{\mathrm{Bm}} \frac{R(t)}{R(t)+k_{\mathrm{RB}}} B(t) \end{cases} \tag{6}$$

with $m_{\mathrm{A}} = m_{\mathrm{B}}$. Apart from the trivial equilibrium without consumers ($A = 0, B = 0, R = s_{\mathrm{R}}$), we now also have the equilibria

$$A = A^* = \frac{a_{\mathrm{R}}(s_{\mathrm{R}} - R_{\mathrm{A}}^*)}{q_{\mathrm{RA}} m_{\mathrm{A}}}, \;\; B = 0, \;\; R = R_{\mathrm{A}}^* = \frac{k_{\mathrm{RA}} m_{\mathrm{A}}}{f_{\mathrm{Am}} - m_{\mathrm{A}}} \tag{7}$$

($A$ survives, $B$ dies, food level determined by consumption by $A$), and

$$A = 0, \;\; B = B^* = \frac{a_R(s_{\mathrm{R}} - R_{\mathrm{B}}^*)}{q_{\mathrm{RB}} m_B}, \;\; R = R_{\mathrm{B}}^* = \frac{k_{\mathrm{RB}} m_{\mathrm{B}}}{f_{\mathrm{Bm}} - m_{\mathrm{B}}}. \tag{8}$$

($B$ survives, $A$ dies, food level determined by consumption by $B$). Which equilibrium is the stable one in the presence of both consumers depends on the critical resource densities $R_{\mathrm{A}}^*$ and $R_{\mathrm{B}}^*$ only.

***Paradox 3:*** Assuming that the stable resource level $s_{\mathrm{R}}$ is sufficiently high to have positive stationary consumer densities, there are two possibilities. Either $R_{\mathrm{A}}^* < R_{\mathrm{B}}^*$, and species $A$ prevails, or $R_{\mathrm{A}}^* > R_{\mathrm{B}}^*$, in which case eventually $B$ gets the upper hand. The other species dies out. The special case $R_{\mathrm{A}}^* = R_{\mathrm{B}}^*$ will be discussed later. Note that the conversion factor $q$ plays no role here, because it is not contained in the equation for the critical resource densities. Within the model a consumer can use an improved foraging strategy to increase its growth rate in order to outcompete other consumer species foraging for the same resource, even if in the process the conversion factor is increased. If that is the case in a real biological system the final density of the survivor invading in a stable system may even be lower than the original stable density of the one now becoming extinct. It is sufficient that the prevalent species outperforms the lesser one in terms of eating away so much of the food supply, that the other is unable to survive on what remains. ***It is competition for the resource only, what the consumer does with the resource is irrelevant. That contradicts our initial assertion, where we stated the chances of survival could be increased by increasing the equilibrium consumer biomass***, so was that assertion wrong, or is the model wrong, or do we have the wrong perspective?

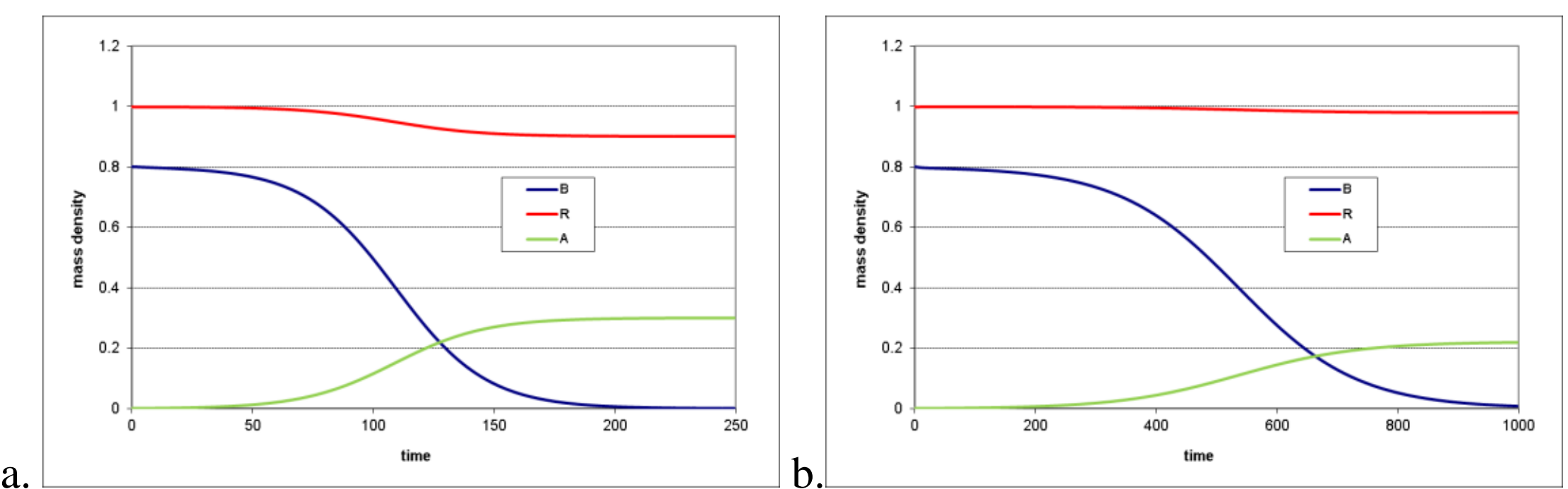

Figure 7. Time development of the densities of two species competing for a single resource. The half saturation constant for $\boldsymbol{A}$ is 10% smaller than that for $\boldsymbol{B}$ (a), the conversion factor $q$ of $\boldsymbol{A}$ is four times that of $\boldsymbol{B}$. All other parameters are the same for $\boldsymbol{A}$ and $\boldsymbol{B}$. Initially species $\boldsymbol{B}$ and the resource are at equilibrium, but eventually $\boldsymbol{A}$ successfully invades and $\boldsymbol{B}$ becomes extinct. Even if the difference in $\boldsymbol{k}_{\mathrm{R}}$ is just 2% (b), still $\boldsymbol{A}$ prevails, but it takes much longer.

In Figure 7 parameters are chosen such that the paradoxical scenario is exactly what happens. For the resource $a_{\mathrm{R}} = 1$ and $s_{\mathrm{R}} = 1.2$. For the consumers $m_{\mathrm{A}} = m_{\mathrm{B}} = 1$, $f_{\mathrm{Am}} = f_{\mathrm{Bm}} = 2$, $k_{\mathrm{RA}} = 0.9$, $k_{\mathrm{RB}} = 1$, $q_{\mathrm{RA}} = 1$, and $q_{\mathrm{RB}} = 0.25$. This means that $A$ is slightly more effective than $B$ n in acquiring the nutrient, growing at a higher rate at the same nutrient density, but $A$ needs much more of the nutrient in order to do so. The equilibrium values are $A^* = 0.3$, $B^* = 0$, and $R^*_{\mathrm{A}} = 0.9$, respectively $A^* = 0$, $B^* = 0.8$, and $R^*_{\mathrm{B}} = 1$. As initial value we chose $A(0) = 0.001$, $B(0) = 0.8$, and $R(0) = 1$, very close to the stationary point (8) of the system, which is a saddle point in the presence of $A$. Indeed after a while the density of both $B$ and $R$ can be seen to drop, while that of $A$ slowly increases. The system moves away from the saddle point and eventually reaches the stable node without presence of $B$ given by eqn (7). Even though species $B$ needs much less food to grow than species $A$, it is the rate at which the species grows that eventually decides the balance. Because of the food requirements for $A$, which is removing only 25% of the stable supply level of the resource, there simply is just not, and only just not enough left for $B$. Even if the difference is much smaller $k_{\mathrm{RA}} = 0.98$ the dynamics is the same, but the takeover is much slower.

There is a paradox in this. Initially the net rate (inflow – outflow) at which food is supplied is at 17% of the maximal rate $a_{\mathrm{R}} s_{\mathrm{R}} = 1.2$; the majority of the inflowing nutrient (83%) leaves the system untouched ($a_{\mathrm{R}} R^*_{\mathrm{B}} = 1$). This is sufficient to maintain a sizeable, stable population of $B$ in the absence of $A$. Eventually the net supply rate is at 25% of its maximum ($a_{\mathrm{R}} R^*_{\mathrm{A}} = 0.9$), as $A$ is eating away more, sustaining a substantially smaller population of $A$ than that of $B$. The food density has barely dropped, implying there is still plenty of food around for $B$ to forage for. In fact if $B$ would eat at the same total rate it did originally, the resource density would drop to 0.7 instead of 0.9. One would think that with a little bit of extra effort both $A$ and $B$ would be able to obtain enough food to happily coexist, without putting a serious dent in the food supply. Not according to the model. Once the food density drops below unity, the growth function for $B$ mercilessly convicts the species to extinction. The growth function depends on the resource density ($R$), not on the resource supply rate ($R'$). It looks like here the paradox is in the original assertion about survival potential and the detailed working of the model. In reality an increase in population size because of behavioural change will most probably increase the survival potential of the species, but the Tilman model shows this may not (always) be true. If any behavioural change results in increased food consumption, even if the population size decreases, within the confines of the model, the newly adopted behaviour will outperform the original one. A smaller population of course is more vulnerable to external fluctuations, but these are not included in the model. In a model with fluctuations the outcome may be more in line with our original expectation. ***The paradox is the result of our expectation based on systems that do have fluctuations, where the model has not, because it is completely deterministic.*** A paradox of a different nature is that an intruder $A$ could invade a system with $B$ living on $R$, voraciously eat away the food, making $B$ disappear, and then the bad food economy of $A$ leaves it vulnerable to extinction as well. Not a paradox at all, just a possible, be it unattractive scenario, of a real ecosystem. It is exactly what happens in the Tilman model if the equilibrium density of $A$ is even lower than in the above example. Within the model the consumer survives, in reality it becomes doubtful whether it will survive in the long run.

**Equal $R^*$**

A special case of the above Tilman model is $R^*_{\mathrm{A}} = R^*_{\mathrm{B}}$. If all parameters are independent this is not very likely to happen, but it may answer the question what happens if part of a population develops a slightly more efficient foraging strategy that allows it to grow with less food. We consider the case that the growth functions for $A$ and $B$ are identical, only the conversion factor of $A$ is slightly less than that of $B$: $q_{\mathrm{RA}} < q_{\mathrm{RB}}$, which means that $A$ consumes slightly less of the nutrient to reach the same growth rate. The answer is disappointingly simple, nothing happens. The reason is that for this particular case the stationary state is not a single point with fully specified values for all three densities, but the whole line connecting the equilibrium points (7) and (8). There are infinitely many equilibria. One of the eigenvalues at the equilibrium points is identical to zero, implying that the dynamics along the associated eigenvector, which is directed along the line, is infinitely slow. Along the line the resource density is fixed at the critical value, and both consumer densities are constant, but not uniquely specified. The equilibrium consumer densities, because the resource density is constant, satisfy

$$q_{\mathrm{RA}}A_{\mathrm{s}} + q_{\mathrm{RB}}B_{\mathrm{s}} = \frac{a_{\mathrm{R}}(s_{\mathrm{R}} - R_{\mathrm{A}}^*)}{m_{\mathrm{A}}} = \frac{a_{\mathrm{R}}(s_{\mathrm{R}} - R_{\mathrm{B}}^*)}{m_{\mathrm{B}}} \quad (9)$$

At which point on the line segment the system will end is determined by the point where it starts. The first two equations in (6) are fully identical, so we have

$$\frac{\mathrm{d}A(t)}{A(t)\mathrm{d}t} - \frac{\mathrm{d}B(t)}{B(t)\mathrm{d}t} = \frac{\mathrm{d}}{\mathrm{d}t}\ln\left(\frac{A(t)}{B(t)}\right) = 0 \quad \Rightarrow \quad A_{\mathrm{s}}B(0) = B_{\mathrm{s}}A(0) \quad (10)$$

This means the ratio between the consumer densities is constant. If $A(0) = 0$, the biomass density of $A$ will stay zero ($A_{\mathrm{S}} = 0$), and $B_{\mathrm{S}} = B^*$, if $B$ is absent initially $A_{\mathrm{s}} = A^*$ and $B_{\mathrm{s}} = 0$. An invading species that just consumes slightly less food than an existing one will remain at a low density if it invades at low density. The total consumer biomass density will increase slightly (fig 8), making the combined population slightly less vulnerable to sudden temporary changes in the food supply, but the “better” foraging behaviour of $A$ does not give it any advantage over $B$.

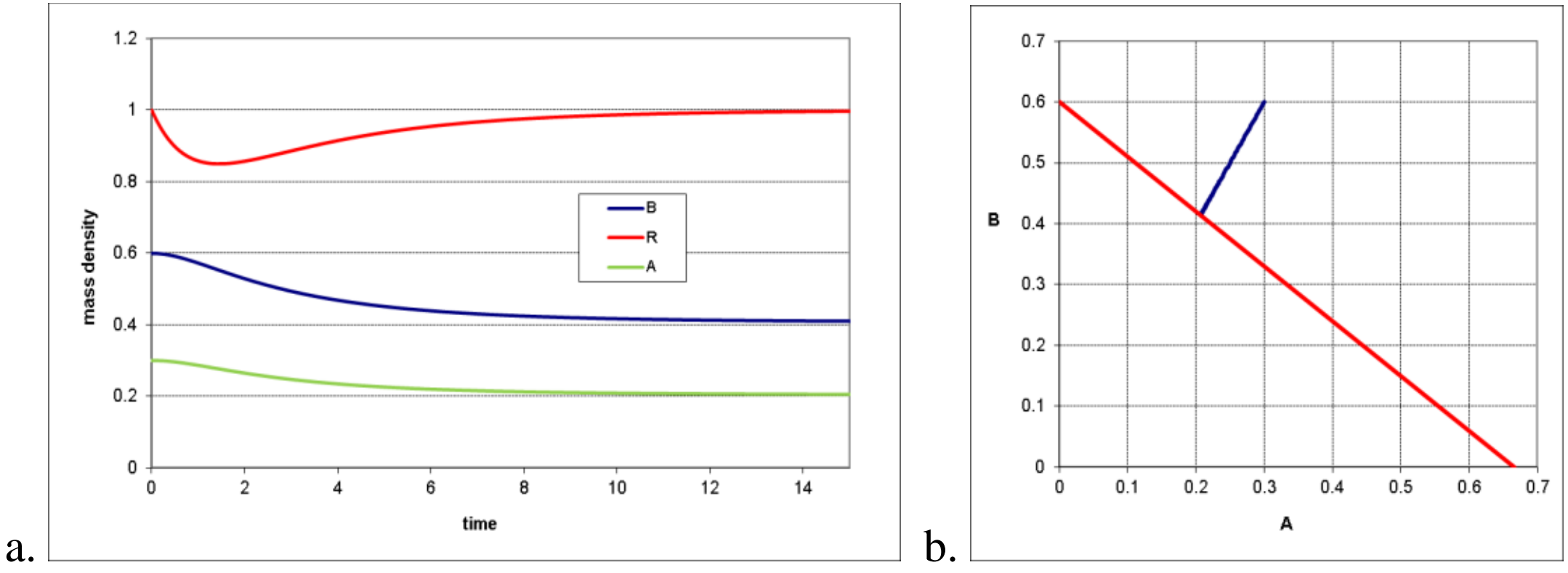

fig 8. Tilman system with two consumer species and one resource. Consumer $B$ is in equilibrium with the nutrient when $A$ invades at high density. For the resource $a_{\mathrm{R}} = 1$ and $s_{\mathrm{R}} =$

1.6. For the consumers $m_{\mathrm{A}} = m_{\mathrm{B}} = 1$, $f_{\mathrm{Am}} = f_{\mathrm{Bm}} = 2$, $k_{\mathrm{RA}} = k_{\mathrm{RB}} = 1$, $q_{\mathrm{RA}} = 0.9$, and $q_{\mathrm{RB}} = 1$. Because with both consumers the food supply is insufficient to sustain the initial mass, the nutrient density drops, but so do the consumer densities. The ratio of the consumer densities stays the same, and in the end the nutrient density veers back to $R^*$, while the sum of the consumer biomass densities is just a little bit higher than the original density of $B$. In the phase plot (b) of the $AB$-plane the trajectory (blue) follows a straight line towards the origin, ending on the line of the equilibria (red).

Within the model only a change in the growth function can accommodate an effect of improved foraging that will lead to an advantage of one strategy above the other. If only a single resource is involved, it will always lead to the consumer with the better strategy taking over the ecosystem, there cannot be sustainable coexistence of two consumer species. That latter conclusion is typical for the model, coexistence within the Tilman model needs at least two resources.

### 2.3. A single consumer depending on two resources.

For a single species depending on two different resources, the dynamical equations are according to Tilman (1982).

$$\begin{cases} \frac{\mathrm{d}B(t)}{\mathrm{d}t} = & f_{\mathrm{B}}(P(t),R(t))B(t) - m_{\mathrm{B}}B(t) \\ \frac{\mathrm{d}P(t)}{\mathrm{d}t} = & a_{\mathrm{P}}(s_{\mathrm{P}} - P(t)) - q_{\mathrm{PB}}f_{\mathrm{B}}(P(t),R(t))B(t) \\ \frac{\mathrm{d}R(t)}{\mathrm{d}t} = & a_{\mathrm{R}}(s_{\mathrm{R}} - R(t)) - q_{\mathrm{RB}}f_{\mathrm{B}}(P(t),R(t))B(t) \end{cases}, \tag{11}$$

Note the different conversion factors for the two resources. Both are consumed, in a fixed proportion. The growth rate of the species depends on both densities but in general only one is limiting. The growth function used is

$$f_{\mathrm{B}}(P,R) = f_{\mathrm{Bm}}\min\left(\frac{P}{P+k_{\mathrm{PB}}},\frac{R}{R+k_{\mathrm{RB}}}\right), \tag{12}$$

implying that for a given growth rate both resources need to have a certain minimal level, the presence of either resource is essential for growth and stability of the consumer population density. Both resources have a fixed replenishing rate, different for different resources, and independent of the other resource. Also the dilution rates are in principle different, but fixed. If there is a physical background, such as a river delivering the nutrients to a lake, the dilution rates may be identical. Before we start investigating what may happen in such a system, we first recapitulate some of the mathematics.

There are two stationary states. One is the trivial equilibrium ($B = 0, P = s_{\mathrm{P}}, R = s_{\mathrm{R}}$) without consumers and both resources at their replenishment level. The nontrivial stationary point can be found as the intersection of two curves in the $PR$-plane (see Van Opheusden, 2015 for a more detailed explanation). One curve is the zero growth isocline for $B$ (where $\mathrm{d}B(t)/\mathrm{d}t = 0$), two half lines, parallel to the $P$-axis and $R$-axis, and starting in $(P^*, R^*)$

$$\left(P = P *= \frac{k_{\mathrm{PB}}m_{\mathrm{B}}}{f_{\mathrm{Bm}}-m_{\mathrm{B}}}, R > R^*\right),\quad \left(P > P^*, R = R^* = \frac{k_{\mathrm{RB}}m_{\mathrm{B}}}{f_{\mathrm{Bm}}-m_{\mathrm{B}}}\right). \tag{13}$$

The second curve in this case is a straight line

$$\frac{a_{\mathrm{P}}}{q_{\mathrm{PB}}}(s_{\mathrm{P}} - P) = \frac{a_{\mathrm{R}}}{q_{\mathrm{RB}}}(s_{\mathrm{R}} - R). \tag{14}$$

through the so-called supply point ($P = s_{\mathrm{P}}$, $R = s_{\mathrm{R}}$). At the intersection not only $\mathrm{d}B(t)/\mathrm{d}t = 0$, but also $\mathrm{d}P(t)/\mathrm{d}t = 0$ and $\mathrm{d}R(t)/\mathrm{d}t = 0$, so we have equilibrium. The exact location of the intersection lies on one of the two parts of the isocline for $B$ (see also Fig. 9b). The stationary point $(B, P_{\mathrm{B}}, R_{\mathrm{B}})$ is one of the following two

$$B = B^* = \frac{a_{\mathrm{R}}(s_{\mathrm{R}}-R^*)}{q_{\mathrm{RB}}m_{\mathrm{B}}},\ P_{\mathrm{B}} > P^*,\ R_{\mathrm{B}} = R^* \tag{15a}$$

$$B = B^* = \frac{a_{\mathrm{P}}(s_{\mathrm{P}}-P^*)}{q_{\mathrm{PB}}m_{\mathrm{B}}},\ P_{\mathrm{B}} = P^*,\ R_{\mathrm{B}} > R^* \tag{15b}$$

The stationary consumer population is positive, if $s_P > P^*$ and $s_R > R^*$. That means that the supply point lies between the two parts of the $B$-isocline. If such is the case, the stationary point is a stable node or stable vortex, while the trivial equilibrium is a saddle point. If not, the trivial equilibrium is a stable node, leading to extinction.

***Paradox 4:*** Now that we have the mathematical background covered, we can turn to the ecology. We know that both nutrients are essential, but in general only one will be consumed to the point that it reaches it minimal level to sustain the population (eqn 15). This is called the limiting nutrient. What happens with the other, and does the fact that its density is not minimal open options for other foraging strategies? We choose a system with parameter values $a_P = a_R = 1$, $s_P = s_R = 1$ (the two nutrients have the same parameters), $m_B = 1$, $f_{Bm} = 3$ (the maximal intake rate is substantially larger than the mortality), $k_{PB} = 0.8$, $k_{RB} = 1.2$ (the intake rate for $P$ increases steeper than that of $R$), $q_{PB} = 1.2$, and $q_{RB} = 0.6$ (the consumer eats more of $P$ than of $R$). This implies that the minimal required levels of the nutrient are $P^* = 0.4$ and $R^* = 0.6$, and the equilibrium is at $P_B = 0.4$, $R_B = 0.7$, and $B^* = 0.5$. We start at $B(0) = 0.001, P(0) = 0, R(0) = 0$. Figure 9 shows the time development of the three densities (Fig.9a) and the phase plot (Fig.9b) indicates how the trajectory navigates between the stationary points. Note that the phase plot only shows the $PR$ -plane, the consumer density $B$ is not shown. In the present case the coexistence point lies on the half line of constant $P$, this nutrient is brought down to the lowest level that can still support the consumer. The density of $R$ remains above the minimum value that compensates for the innate mortality of the consumer. The reason is the relatively low conversion factor for $R$ ($q_{RB}$), or inversely, the relatively high associated yield factor. Because the consumer eats $P$ at a higher rate than $R$, the intake ratio $q_{RB}/q_{PB}$ and thus the slope of the dashed line in fig 9 is such that we end up in the equilibrium given by eqn 15b

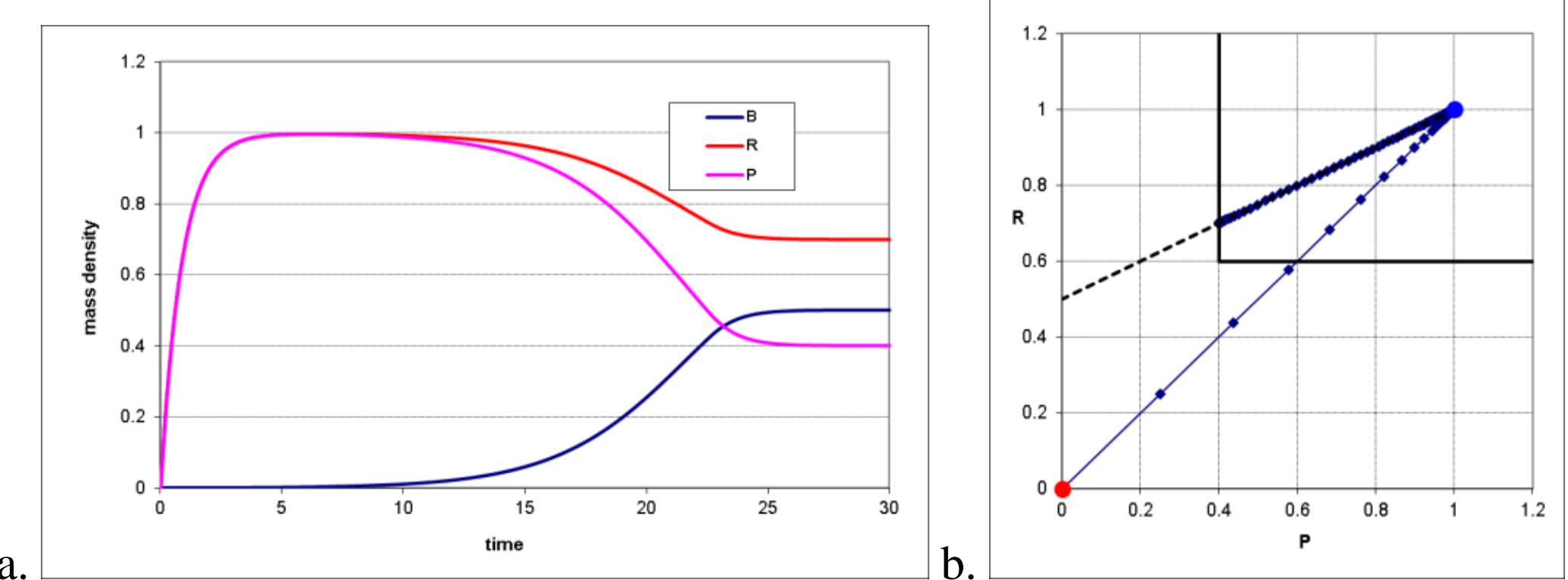

Figure 9. Dynamics of a single consumer depending on two essential resources. Parameter values are $\boldsymbol{a_{\mathrm{P}}} = \boldsymbol{a_{\mathrm{R}}} = \mathbf{1}$, $\boldsymbol{s_{\mathrm{P}}} = \boldsymbol{s_{\mathrm{R}}} = \mathbf{1}$, $\boldsymbol{m_{\mathrm{B}}} = \mathbf{1}$, $\boldsymbol{f_{\mathrm{Bm}}} = \mathbf{3}$, $\boldsymbol{k_{\mathrm{PB}}} = \mathbf{0.8}$, $\boldsymbol{k_{\mathrm{RB}}} = \mathbf{1.2}$, $\boldsymbol{q_{\mathrm{PB}}} = \mathbf{1.2}$, and $\boldsymbol{q_{\mathrm{RB}}} = \mathbf{0.6}$. The time plot (a) shows that initially the two resource densities increase to their stable levels, while the consumer remains at low density. Once the consumer density starts increasing, the resource densities drop and the stable coexistence point is approached. The phase plot (b) in the $\boldsymbol{PR}$ -plane shows the same behaviour. The small dots are the equidistant time points of the numerical integration. Starting from the initial point close to the trivial equilibrium (red dot) it rapidly moves to the supply point (blue dot), which lies inside the quadrant defined by the $\boldsymbol{B}$-null cline (thick black half lines). From there it moves along the dotted line as defined by equation (14) to the stable coexistence point. The slope of the dotted line is determined by the intake ratio.

It looks like we have another paradox here. In order to survive, according to (11) a minimal concentration of $R^*$ is needed, and indeed the actual concentration $R_{\mathrm{B}}$ at equilibrium is larger, as expected. ***But with more of R around, the consumption of this nutrient is increased beyond what is actually needed, so why would the consumer take the luxury of eating more of R than the minimal requirement,*** what is the use of this luxury consumption? The answer is in the chemostat equation paradox. When the density of $R$ is higher than $R^*$, the consumption rate of $R$ is actually lower than the one at the critical density. The additional flow is not consumed, but leaves the system untouched. It is all about the growth rate, as expressed by (12), and that does not depend on the consumption rate, but on the density of the limiting nutrient $P$. An increase in $R$ has no effect on the growth function (because of the 'min' function in eqn 12) , and hence the uptake of $R$ by the consumer (the last term in eqn 11) is independent of the density of $R$. The fact that for a single resource as in eqs 1 & 2 the uptake does depend on $R$ can also be the source of the paradox. When $P$ is the limiting resource, the density of $R$ drops out of the equation. There is no such thing here as "luxury" consumption of $R$, unless one would want to use that for any consumption of that resource.

If we take $q_{\mathrm{RB}} = 0$, the consumption of that resource is zero and its density reaches its supply level. In this case the full $B$ population survives on $P$ only, and more importantly, at the same population size. Any effort invested in foraging for $R$ seems wasted from this perspective. Obviously that is not what the model is intended to describe. Both resources are needed and both need to be present at a certain level. Resources are consumed at a fixed ratio until one of them reaches that level. At that point the other resource is at a higher level than minimally needed, because it is consumed at a lower rate. One may expect that a higher concentration may lead to a higher consumption, but such is not the case for the current model, in particular the growth function (eqn 12), which is independent of $R$ if $P$ is limiting (and vice versa). The paradox is resolved by realising that the surplus of resource $R$ is not consumed, but wasted. ***There is no luxury consumption, the consumer has the "luxury" to let some food pass by untouched.*** The consumption of the other resource could, in principle be increased further to alter the model parameter such that an increased growth potential is obtained. That raises the question whether this "wasted" nutrient flow might be used by the consumer.

The actual question thus is not why the consumer is eating more of the non-limiting resource than needed, but whether the consumer could change its foraging strategy for the two resources to reach the same growth rate with less effort. This would open up possibilities to improve chances of survival with the same effort. Within the model there is some room. The density $P^*$ is limiting growth, the consumer eats at 60% of the maximal provision rate, and cannot eat more of $P$ without having the density of that resource drop below $P^*$. The density $R^*$ is not limiting, a consumer may eat more of this resource and use the energy gain to forage more efficiently for $P$. Note that we do not mean that the consumer might use the additional $R$ to substitute for $P$, we still assume that in the growth function (12) both are essential, and the isocline is unaltered. We assume that an additional intake of $R$ in some way can lead to a better food economy for $P$. Indeed if we drop the conversion factor of $P$ to $q_{\mathrm{PB}} = 0.9$, we find stationary densities $P^* = P_{\mathrm{B}}$, $R^* = R_{\mathrm{B}}$, and $B^* = 0.67$. At this point the consumer population still uses the same 60% of the maximal rate for $P$, and 40% of that for $R$ instead of the original 20%. So can we conclude that it helps to employ the additional food source $R$ to improve the growth potential? Not really. This is simply one numerical example where the strategy does seem to be effective, a 25% extra food intake gives 33% more consumers. That sounds like a good deal indeed. But is it fair? Wouldn't an increase in intake of $R$ lead to a drop in efficiency, and is the decrease in the conversion factor of $P$ realistic?

The answer is in equation (15). If the coexistence point is along the $P = P^*$ half line, the stationary resource density is fully determined by the consumption parameter values for this resource. A change in for instance the conversion rate of $R$, with other parameters fixed, is simply compensated by a different resulting value for $R_{\mathrm{B}}$, to give the same $B^*$. This effect can also be observed in Fig.9b, a change in the slope of the dotted line gives a change in $R_{\mathrm{B}}$, as long as the intersection point stays above $R^*$. If an increase in $q_{\mathrm{RB}}$ can be used to lower $q_{\mathrm{PB}}$, no matter how small the decrease, the total effect, according to eqn (15b), will always be an increase in $B^*$. The optimum is reached when both resources are consumed to such a level that both are restrictive for growth. In this case the intersection point of the half lines is the stationary point. How large the effect is cannot be decided within this model, one would need an additional model describing the actual foraging process.

## 3. Conclusions and Discussion

For the simple snail – lettuce model in the introduction we have shown that an extended stability analysis, including all stationary states of the system, as well as the rates at which these states are approached or departed, can help avoid misunderstandings caused by just looking at the asymptotically stable state. Before this asymptotic state is reached the system may pass along other states where it may dwell for quite a while before moving on. Not only are the unstable stationary states possibly important, also the time it takes to move from one state to the other sets the stage for what we may expect for the system to do, without the need to perform a full numerical simulation and follow the detailed trajectories. There is no reason to call for strange attractors or limit cycles or other phenomena from non-linear dynamics of complex systems, simply two saddle points and a stable vortex suffice to lead someone up the garden path.

Within the Tilman model we have looked for ways how a consumer could deal with the available resources, for which we have loosely used the term "foraging", and what that would imply to the parameters and the resulting dynamics within the confines of the model. In several cases that leads to (constructed) paradoxes, the unravelling of which we hope adds to the understanding of (the working of) the model.

A single consumer species using a single resource has a variety of options to increase the consumption to the level where all available food is being consumed. That leads to a paradox, because if there is no food left, it becomes hard to find any. In practice the stationary food level will never become exactly zero, it simply will be lower the better the consumer is geared to quickly remove all incoming food. Only if all inflowing food is eaten instantaneously will the equilibrium density become zero, but in that extreme case one had better look at the consumer population itself as the ecosystem in which the food is injected, and see how it is digested within the body of the consumer. The optimal strategy in this case is indeed an optimum in the mathematical sense, it can be approached as closely as one would want, but it can never be reached in this model. Of course it is possible to design models in which it is assumed that all food is being consumed and the detailed dynamics of the food concentration is left out of the model (see e.g. Huston & Smith, 1987; Otto & Day, 2007), and leaving the competition for the resource to parameters that just apply to different consumer species in such models.

The paradox we discussed is associated with the food supply dynamics, according to the chemostat equation. Further analysis showed how the paradox can be resolved, but it is a valid question whether this kind of dynamics applies to any real ecosystem at hand. There already has been a discussion of the validity of the Tilman model, and Berendse (1994) developed another model based on the experimental results of de Wit (1960). More recently, Craine (2005) compared the theories of Tilman and Grime, which resulted in a debate about plant strategy theories (see Grime 2007: Tilman 2007 and Craine 2007). The result is that Craine suggests to make a synthesis of the Tilman and Grime theories, while trying to understand the mechanisms behind plant coexistence. Note that the Tilman model was developed originally for algae (Tilman, 1977) and later the concept is transferred to plants (Tilman, 1982), and the resources were assumed to be abiotic compounds, such as phosphorous or nitrogen. For such an ecosystem the notion of foraging is much less clear than for snails eating lettuce. While snails may decide to switch to cabbage if the lettuce runs out, plants cannot simply decide to consume silicium dioxide once carbon dioxide becomes scarce (though it could be a tremendous solution at high altitudes). That does not mean plants do not have ways to adapt to their environment. On the contrary, they just do it in different ways than animals, and with much success. We have steered clear from making explicit what actual consumer behaviour is responsible for what specific parameter change in the model, and restricted the discussion to a parameter study of the model, with a clear connection to what parameters could be open for improvement.

For a Tilman system of two consumer species and a single nutrient we have seen that it may take a long time before the "better" one fully takes over. Moreover, if the consumption parameters of the different consumers are such that their critical resource densities are identical, even if the total food economy of the consumers is different, the model allows these to coexist at a fixed density ratio. The conclusion can be extended to a system with any number of consumers and nutrients. For larger systems also more complex, chaotic dynamics is possible (see for example Huisman & Weissing, 1999) but the same more simple dynamics of the smaller systems can already accommodate the existence, be it maybe for a finite but long time period, of any number of consumers. This forms a possible explanation of the so-called plankton paradox of the Tilman model: how can many species of organisms coexist in the plankton, even if they are competing for the same limited number basic nutrients? It is possible that such a system, when left alone, in the long run would see the disappearance of the majority of the species, but in practice the system is not left alone and undergoes major disturbances on time scales (much) smaller than what is needed for the slightly less performing species to become extinct (see also Grime 2007).

It may seem paradoxical that the better consumer, the one outcompeting the other, does not need to have a better food economy. The notion of a food economy and a quality that can be contributed to it is an intuitive one from our general experience, not an integral part of the model. Within the model it suffices to be able to survive at a lower critical resource density than the competition. It isn't even necessary to eat all the available food, just bring the level down to below what the others need. In that sense it is maybe more important to realise that the bad food economy is actually an asset, to remove the competition it suffices to take away the food, so if winning is the name of the game, this is your chance of cheating. It is not what the model is intended for, obviously, but it may describe what happens if invasive species outperform native ones, only to set the stage for their own demise once the circumstances change. Considering such competition and coexistence Huston and DeAngelis (1994) develop a model where the perfect mixing of nutrients does not hold. Plants will affect only the nutrient concentration in their immediate surroundings, and thus compete only with their immediate neighbours. In the limit of low resource input and high transport through the environment their model shows similar dynamics as models assuming complete mixing.

A single consumer of two essential resources consumes these within the model in a fixed ratio. The effect is that in practice always one of the resources will be limiting, and the limiting resource will be consumed to its limiting density. The non-limiting resource in this case will be consumed at a rate that is determined by the availability of the limiting resource and the ratio of the consumption rates. Its density at that point is higher than needed, implying that the excess is going to waste; it isn't used by the consumers because they cannot use it, since there is not enough of the limiting resource they would need in the process. The reverse of course is true, if the consumer can find ways of using the non-limiting resource to its advantage, it can only do so to the point where it does become the limiting one. From that point on the focus of any further improvements shifts to the other resource, or to both resources combined.

We have presented several paradoxes for specific ecological models and showed how they can be avoided or resolved. The origin of the paradoxes is diverse, and so is the pathway leading to their resolution. In general a different perspective is needed, the problem what perspective is not always clear, and such a perspective may not always be available, at least not for the paradox at hand and the person perceiving the paradox. If such is the case the remaining option, for the moment, is to accept the paradox. We have provided pathways for the specific paradoxes we presented that we hope allow the reader to resolve these, and provide strategies to resolve similar paradoxes in more general cases.

**Acknowledgements**

We like to thank Wopke van der Werf for comments on an earlier version of this work, and Hans Metz who gave a critical review of the manuscript.

**Statements and Declarations**

This research did not receive any specific grant from funding agencies in the public, commercial, or not-for-profit sectors. The authors declare that they have no known competing financial interests or personal relationships that could have appeared to influence the work reported in this paper (so no conflict of interest). Moreover, there are no data pertaining to this research.

## Appendix. The Snail-Lettuce Model

A farmer grows lettuce, which is eaten by snails. In a very simple model the ecosystem represented by the biomass $L(t)$ of lettuce and snails $S(t)$ is modelled as

$$\begin{cases} \frac{\mathrm{d}L(t)}{\mathrm{d}t} = & aL(t)(1 - L(t)/x) - bL(t)S(t) \\ \frac{\mathrm{d}S(t)}{\mathrm{d}t} = & -cS(t) + dL(t)S(t) \end{cases} \qquad \text{(A0)}$$

with $a$, $b$, $c$, and $d$ positive constants. In the absence of snails, the lettuce in the model will grow at an initial rate $a$, until it reaches a density $x$ (the carrying capacity). In the absence of lettuce, the snails will die in an exponential relaxation process with rate constant $c$. If both are present the snails eat the lettuce at a rate proportional with both densities and proportionality constant $b$, with the lettuce biomass turned into snail biomass, as given by the constant $d$. The system will then tend to an equilibrium with densities $L_s = c/d$, $S_s = (1 - c/dx)\ a/b$. We assume $x >> c/d$ , that is, the carrying capacity largely exceeds the equilibrium density of lettuce coexisting with snails. In equilibrium any increase in lettuce biomass is eaten away by the snails, and any increase in snail biomass is nullified by snails dying. In equilibrium the farmer is producing dead snails rather than lettuce.

### Effect of farmer interaction

Assuming this is not what the farmer had in mind, he or she has two options to change the equilibrium state: (1) kill off the snails, or (2) speed up the growth of the lettuce. The first option amounts to an increase of the parameter $c$, the mortality rate of the snails. The second option amounts to an increase of the parameter $a$, the growth rate of the lettuce. The parameters $b$ and $d$ describe the interaction between the snails and the lettuce, which is no business of the farmer. Because of the (changed) interference of the farmer with the ecosystem, the equilibrium shifts. Now we find a paradox. If the mortality rate $c$ of the snails is increased, it is not the snail density at equilibrium that is heavily affected, but rather the lettuce density at equilibrium that increases.

If on the other hand the growth rate *a* of the lettuce is increased, the equilibrium lettuce density is fully unaltered, instead the snail density goes up. This is counterintuitive, killing off snails should bring the snail density down. Similarly boosting lettuce growth should increase lettuce density. Indeed that is exactly what happens, right after the parameter is changed, but after a while the effect is counteracted by the response from the interacting snail-lettuce system. An increase in lettuce density does not go unnoticed by the snails (fig A1), their density increases as well, until an equilibrium is reached in which again all lettuce growth is converted into dead snails. More dead snails, because their mortality rate is much higher, while the equilibrium density is only slightly reduced. A decrease in snail density leads to an increase in lettuce (fig A2), and again the snails seize the opportunity. Lettuce density goes down again until the same equilibrium value for the lettuce is reached as before, but a higher snail density and a similarly increased rate of dead snail production.

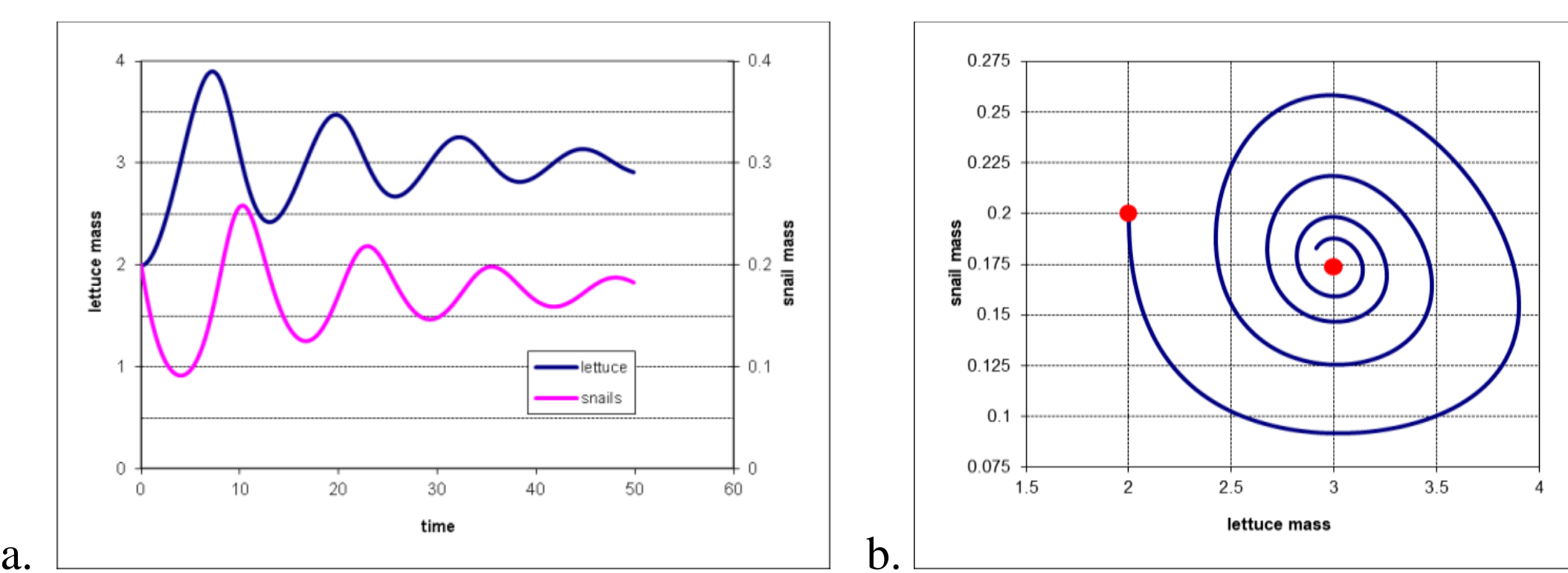

fig. A1. Effect of an increase in the death rate *c* of the snails in the snail-lettuce model. The system is at equilibrium at parameter values $a = 0.4$, $b = 1.6$, $c = 0.6$, $d = 0.3$, $x = 10$ (so $S_s = 0.2$ and $L_s = 2$), when *c* is increased to 0.9. On the short term the snail mass *S* (cyan line in fig a) goes down as expected, while the lettuce mass *L* goes up (blue line, note the difference in scale of the lettuce and snail mass). On the long term the effect is considerably smaller; the lettuce mass does go up, the snail mass goes down only slightly. The phase diagram (fig b) shows how

the system spirals from the original equilibrium value to the new one (red dots), with an increased lettuce mass (horizontal axis) and almost the same snail mass (vertical axis).

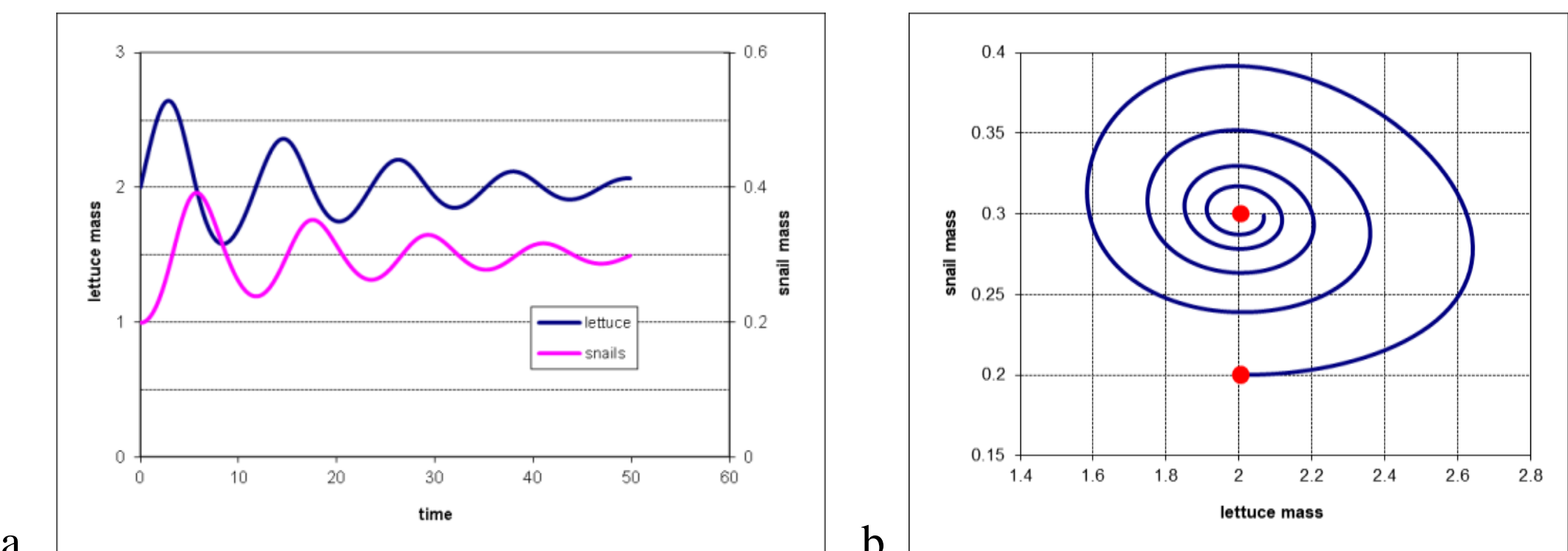

fig A2. Effect of an increase in growth rate $a$ of the lettuce. The system is at equilibrium at parameter values $a = 0.4$, $b = 1.6$, $c = 0.6$, $d = 0.3$, $x = 10$ (so $S_s = 0.2$ and $L_s = 2$), when $a$ is increased to 0.6. On the short term the lettuce mass $L$ (blue line in fig a) indeed goes up, but so does the snail mass $S$ (cyan line). On the long term the lettuce mass returns to the original equilibrium value, the snail mass goes up. The phase diagram (fig b) shows how the system moves from the original equilibrium value to the new one, with the same lettuce mass and an increased snail mass.

**Resolving the paradox**

In fact there is no contradiction, it only looks like there is one, so we have a paradox indeed. Our notion of the system is too simple. The system does react instantaneously as we expect it to do, but in the long run it does quite the opposite, and the effect can be explained and understood by looking at the detailed dynamics of the system. There is not just one equilibrium, in fact the system has three (red dots in fig A3b). One is the trivial equilibrium ($T$) at which there is no lettuce and no snails. This is a saddle point; if there are only snails, the snails die. A second equilibrium is where there is only lettuce ($L$). Also this is a saddle point; the lettuce grows until it reaches the carrying capacity. The coexistence equilibrium ($C$) as has been dealt with above, for the parameter values as used in the model calculations, is a stable vortex; in the phase plot (fig A3b) the system spirals towards it. In the current system the saddle point $L$ is the most relevant equilibrium for the farmer growing lettuce, even though it is an unstable one. Starting from a low snail and lettuce density, that is close to the unstable trivial equilibrium $T$, the system moves away along the unstable manifold (only lettuce) towards the saddle point $L$. Because there are still a few snails, they eventually start multiplying and eating a sizable portion of the lettuce, so the system moves away from the second saddle point as well, starting a series of loops with alternating high and low snail and lettuce density that will ultimately bring it to the stable coexistence equilibrium $C$. Any sensible farmer growing lettuce will not wait for this, but harvest the lettuce right before the snails claim their share. Better avoid the snails than attack them.

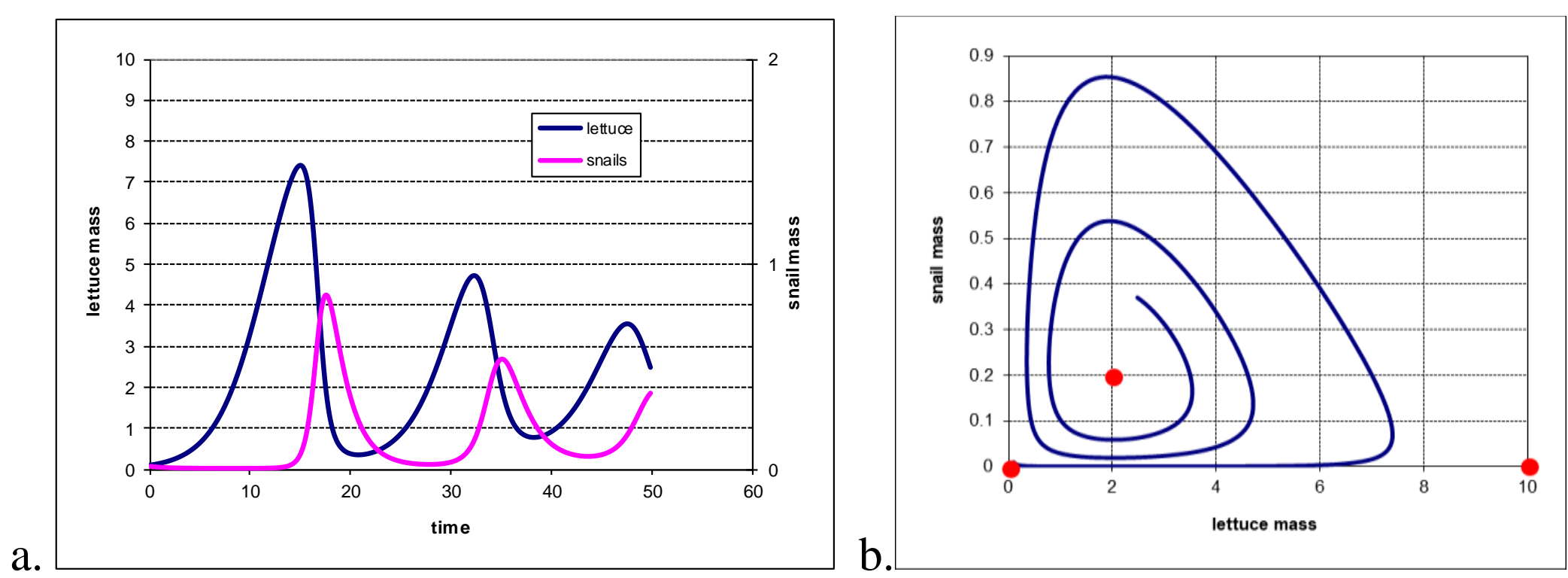

fig A3. Dynamics of the snail-lettuce system starting at low lettuce $(L(0) = 0.1)$ and snail mass $(S(0) = 0.01)$. Parameter values are $a = 0.4$, $b = 1.6$, $c = 0.6$, $d = 0.3$, $x = 10$ (so $S_s = 0.2$ and $L_s = 2$). The lettuce mass grows rapidly (a), while the snail mass remains very low. Once the snails start multiplying seriously, they eat away most of the lettuce, snails die, and the cycle repeats. Eventually the system will reach the stable equilibrium. In the phase plot (b) the system moves away from the trivial equilibrium $T$ at (0,0), towards the saddle point $L$ at (10,0), and bends off to spiral towards the stable coexistence point $C$ at (2,0.2).

The full equilibrium properties of the system can be derived from the stability analysis below. This analysis not only provides the equilibria as described above, but also the stability properties of those equilibria, including the rates at which the system moves close to the equilibria. These rates in turn determine the relevant times scales of the system. Below we perform the full mathematical stability analysis of this system, we now proceed with the conclusions of it. In this case the time scale at which the lettuce is harvested by the farmer is about 10 time units, related to the rate at which the system moves away from the trivial equilibrium ($\tau$ = 2.5 units). The time scale at which the system relaxes towards the final equilibrium is 25 units, implying that it takes of the order of 50-100 time units to really come close to equilibrium. One may check the time plots, the point is that the stability analysis suffices, there is no need to actually solve the equations (cf make a numerical approximation as has been done). The relevant time scale in this case is the short one, the relevant equilibrium an unstable one. The cause of the paradox, from the perspective of the farmer growing lettuce, is looking at the wrong time scale. For a farmer growing snails, a different perspective gives a different picture, of course, and similarly for the snails, the lettuce and the theoretical biologist.

**Conclusion**

This is an example of how a relatively simple mathematical model, and a simple model analysis, can lead to a paradox. Just looking at the stable asymptotic equilibrium state of the system tells us little about the dynamical behaviour away, and towards that equilibrium. Apart from the equilibrium itself, which can relatively easily be found and which shows how changes in the input of the model affect the output, also the rate at which the equilibrium is approached is an important consideration. If that rate is very low, and consequently the time scale (the inverse of the rate) at which equilibrium is reached is very large compared to the time scales of changes in the system parameters, the equilibrium may never be reached and other, possibly unstable equilibria are more relevant. Important is that the same mathematical procedure that gives us the equilibria and their stability properties, can also produce these time scales, at least near the equilibrium.

**Mathematical Stability Analysis**

The biomass $L(t)$ of lettuce and snails $S(t)$ is modelled as

$$\begin{cases} \frac{\mathrm{d}L(t)}{\mathrm{d}t} = & aL(t)(1 - L(t)/x) - bL(t)S(t) \\ \frac{\mathrm{d}S(t)}{\mathrm{d}t} = & -cS(t) + dL(t)S(t) \end{cases} \tag{A1}$$

This is a nonlinear homogeneous system of differential equations, which we write as

$$\begin{cases} \frac{\mathrm{d}L(t)}{\mathrm{d}t} = & f(L(t), S(t)) = & aL(t)(1 - L(t)/x) - bL(t)S(t) \\ \frac{\mathrm{d}S(t)}{\mathrm{d}t} = & g(L(t), S(t)) = & -cS(t) + dL(t)S(t) \end{cases} \tag{A2}$$

to prepare for a stability analysis. The equilibria of the system are the solutions of the nonlinear system of algebraic equations

$$\begin{cases} f(L, S) = & aL(1 - L/x) - bLS = 0 \\ g(L, S) = & -cS + dLS = 0 \end{cases} \tag{A3}$$

This system has three solutions

$(L_0 = 0,\ S_0 = 0),\ (L_1 = x, S_1 = 0),\ (L_2 = c/d,\ \ S_2 = (1 - c/(d\ x))\ (a/b))$.

The stability properties of the equilibria are found by looking at the Jacobi matrix

$$J(L, S) = \begin{pmatrix} \mathrm{d}f/\mathrm{d}L & \mathrm{d}f/\mathrm{d}S \\ \mathrm{d}g/\mathrm{d}L & \mathrm{d}g/\mathrm{d}S \end{pmatrix} = \begin{pmatrix} a(1 - 2L/x) - bS & -bL \\ dS & -c + dL \end{pmatrix} \tag{A4}$$

For the first equilibrium the Jacobi matrix is

$$J(0, 0) = \begin{pmatrix} a & 0 \\ 0 & -c \end{pmatrix} \tag{A5}$$

with eigenvalues $a$ and $-c$. Because all parameters are positive constants, this means the equilibrium is a saddle point (one eigenvalue positive and one negative). The eigenvalues are the rates at which the solution moves away from or towards the equilibrium. In the present case it implies that if there are no snails, the lettuce mass close to the zero point increases exponentially at a rate $a$, at a time scale $\tau_{0,1} = 1/a$. Similarly if there is no lettuce, the snail mass decreases exponentially at a rate $c$, at a time scale $\tau_{0,2} = 1/c$. In the analysis above we have used parameter values $a$ = 0.4, $b$ = 1.6, $c$ = 0.6, $d$ = 0.3, $x$ = 10; the corresponding time scales are $\tau_{0,1} = 2.5$ and $\tau_{0,2} = 1.67$.

The Jacobi matrix for the second equilibrium is

$$J(x,0) = \begin{pmatrix} -a & -bx \\ 0 & -c+dx \end{pmatrix} \tag{A6}$$

with eigenvalues $-a$ and $dx - c$. The first eigenvalue is always negative. In the absence of snails the lettuce mass close to the equilibrium behaves as

$$L(t) = x + (L(0) - x)\ \exp(-at) \tag{A7}$$

and again the rate $a$ defines a time scale $\tau_{1,1} = 1/a$. The second eigenvalues can be either positive or negative, depending on the parameter values. For the values as chosen above $dx - c = 2.4$, meaning that the equilibrium is a saddle point. If the snail mass is nonzero, the death rate of the snails is more than compensated for by the growth rate due to lettuce consumption, and the snail mass increases exponentially at this rate, with time scale $\tau_{1,2} = 0.42$. Because the lettuce mass decreases proportionally, the system moves away at a tangent, along the corresponding eigenvector (see fig A3b).

The Jacobi matrix for the third equilibrium for the given parameter values becomes

$$J(2,0.2) = \begin{pmatrix} -0.08 & -3.2 \\ 0.06 & 0 \end{pmatrix} \tag{A8}$$

with complex eigenvalues $-0.04 \pm 0.44$i. This implies the equilibrium is a stable vortex, solutions spiral towards the equilibrium point with a period $\tau_{2,1} = 2\pi/0.44 = 14.4$at a rate 0.04, with a time scale $\tau_{2,0} = 1/0.04 = 25$. The value of the period can be checked directly in the time plots, the relaxation towards equilibrium should be compared with the half time, which is more difficult to read from the plot itself, but has the correct order of magnitude.

Important is that it takes the system many periods before it comes really close to the stable equilibrium, so the relevant equilibria are the two saddles and the relevant part of phase space is the line connecting them. Provided there are not too many snails to start with, they can readily be removed from the equation, and the remaining model is just the logistic equation for the lettuce

$$\frac{\mathrm{d}L(t)}{\mathrm{d}t} = aL(1 - L/x) \tag{A9}$$

which can be solved fully analytically.